\documentstyle[11pt]{article}
\input{epsfig.sty}

\def\btau{\mbox{\boldmath $\tau$}}
\def\bpi{\mbox{\boldmath $\pi$}}

\newcommand{\be}{\begin{equation}}
\newcommand{\ee}{\end{equation}}
\newcommand{\ba}{\begin{array}}
\newcommand{\ea}{\end{array}}
\newcommand{\bL}{{\cal L}}
\newcommand{\br}{{\bf r}}

\newcommand{\y}{\'\i} 
\newcommand{\vs}{\vspace{1cm}}

\begin{document}

\begin{titlepage}

\title{
{\bf The Skyrmion Model and the Dynamical Breakdown of Chiral Symmetry } }
\author{
  {\bf F. L. Braghin}\thanks{braghin@if.usp.br} 
\\
{\normalsize Instituto de F\'\i sica da Universidade de S\~ao Paulo} \\
{\normalsize C.P. 66.318,  C.E.P. 05315-970, S\~ao Paulo, SP,     
Brazil }
\\
{\bf I. P. Cavalcante}\thanks{ipc@dfi.ufms.br}
\\
{\normalsize Depto. de F\y sica, CCET, Universidade Federal de
Mato Grosso do Sul,}\\
{\normalsize C.P. 549, C.E.P. 79070-900, Campo Grande, MS, Brazil }
}
\date{}
\maketitle
\begin{abstract}
A scalar field, $\eta({\bf r})$, 
is coupled to the skyrmion.
Its classical value in the vacuum (``condensate") reduces to the 
pion decay constant, $f_{\pi}$, being thus proportional to the chiral
condensate $<\bar{q} q>$. 
A quadratic coupling of the scalar field to the pion kinetic Lagrangian term,
reminiscent from the linear sigma model, is considered.
Its mass is an additional free parameter in the model 
 associated to the lightest scalar isoscalar meson.
Solutions for the two resulting coupled differential equations
are found in several cases. 
The nucleon, as a topological soliton, 
``digs a hole'' in the vacuum making the value 
of the scalar condensate to be close to zero inside the skyrmion.
Some of the observables, as for example the baryon
 masses, may have their values closer to the experimental ones.
Derrick's stability analysis is applied to the model.
\end{abstract}



IF-USP: 2000/2003.

\end{titlepage}


\section{ Introduction}

The description of processes involving strong interactions (mainly at low
and intermediate energies) is very difficult in the frame of the
 Quantum Chromodynamics (QCD) due to its 
non abelian color and flavor structures and strong coupling constants.
Thus effective models, like the Skyrme chiral soliton model (Skyrmion model),
are constructed in such a way as to respect general properties 
from the more fundamental theory, such as chiral symmetry and
its spontaneous breakdown \cite{WEINBERG-THEO}.

The topological soliton model was formulated by Skyrme decades ago
\cite{SKYRME} possessing a conserved topological charge identified 
with the baryonic number. 
A deeper discussion about the real 
existence of minimum energy configurations 
in this model is found for example in \cite{schroers}.
The model encompasses chiral symmetry and its spontaneous breakdown
and has further deeper motivation based on the large $N_c$ QCD 
analysis \cite{THOOFT,WITTEN}.
The quantization of the  rotational 
collective coordinates endows the soliton with spin half.
It strengthens the interpretation
of this soliton as a nucleon as long as it is possible to 
obtain  predictions for hadronic properties \cite{ADKINS}. 
An introductory study of the inclusion of quantum and 
thermal effects was done in \cite{QUANTUM}.

Chiral spontaneous symmetry breaking is believed to be realized 
in the Nambu-Goldstone mode: the pion
decay constant, $f_{\pi}$, is different from zero 
and the pion mass is small in the hadronic scale.
$f_{\pi}$ can be seen as a ``sigma condensate'' of 
the linear sigma model.
This sigma classical value is also identified to the
chiral radius which is the constraint for 
the non linear realization of chiral symmetry as well.
The non zero pion mass 
is usually assumed to break the 
chiral invariance of the Lagrangian explicitly, although only slightly. 
The vacuum is expected to be fulfilled with the scalar quark
condensate ($<\bar{q} q>$) which is proportional to the pion 
decay constant.
To the lowest order, the Gell-Mann--Oakes--Renner relation, which
relates QCD to hadronic degrees of freedom, reads:
\be \label{1} \ba{ll}
\displaystyle{ 2 <\bar{q} q> m_q = - f_{\pi}^2 m_{\pi}^2 . }
\ea
\ee
The experimental value of the pion decay constant 
 is nearly $93 \;$ MeV. 
However, in the context of the Skyrmion model, it
may reach very low values (such as $54 \;$ MeV) 
as long as one takes as input the physical values of  nucleon and delta masses. 
This may suggest by itself that $f_{\pi}$ can acquire different values inside
and outside the nucleon.
The constant value of the chiral radius is
 one trivial (homogeneous) solution of the linear sigma model.
We show in this
work that there are other non trivial (non homogeneous) 
solutions in the presence of 
a nucleon, i.e., it
occurs the formation of a ``hole" in the vacuum
due to the presence of the nucleon in the frame of the Skyrmion model.

The coupling to QCD scalar degrees of freedom has already been 
envisaged in the
frame of the Skyrmion model but in different approaches from that 
exploited here.
Schechter and collaborators \cite{GJJS86} and afterwards
Waindzoch and Wambach \cite{WAWA}  
investigated such kind of effect by coupling a scalar field,
representing a gluon degree of freedom.
The value obtained for that ``gluon condensate'' inside the nucleon
was much higher than in the vacuum, thus forming a kind of
``bag''.
This strengthens some qualitative theoretical 
basis of the so called Bag Models
\cite{BAGMODEL}.
In particular, for light glueballs, the bag was found to be
 very shallow \cite{GJJS86}.
These results are in agreement with the stabilizing mechanism 
discussed in \cite{SZV}. 
However the QCD vacuum is very complex and it is expected to contain,
as discussed above, the quark-antiquark scalar condensate
which is a manifestation of the spontaneously broken chiral symmetry. 
Thus a more 
detailed investigation of the contribution of this complex system 
for the structure
of the nucleon is needed as well as the opposite, i.e.,
the effect of the nucleon on the vacuum.
A more complex system was considered in \cite{MEISS-SCHE} where 
a scalar meson field was also included.

We have earlier considered 
the chiral radius in 
the presence of hadrons as a dynamical variable \cite{RTFNB98,CR01}.
In this work we consider a dynamical scalar degree of freedom 
associated with the 
scalar chiral condensate. 
This is effectively implemented by the coupling of the 
Skyrmion to a self interacting scalar field.
As  shown is sections II and III, the  scalar condensate
(``sigma condensate'')
is constrained by the form of the potential which 
endows the model with  the spontaneous symmetry breaking:
in the vacuum it assumes the usual value of the pion decay 
constant, $\eta_{vacuum} = 93 \;$ MeV.
In the next sections we show the Lagrangian density and equations of the
fields of this extended Skyrmion model. 
There are two coupled differential equations: for the soliton profile function,
$F(r)$, and for the scalar isoscalar field, $\eta(r)$.
The numerical solutions are investigated for some cases 
and hadronic observables are calculated. 
We investigate the effect of the magnitude of the
mass of this scalar field - 
the additional free parameter of our model -
privileging values 
between $560 \;$MeV and $980 \;$MeV, which correspond respectively 
to the sigma (eventually the chiral partner of the pion) 
\cite{SIGMA,TORNQVIST,BEDIAGA}
and to the $f_0(980)$ meson \cite{TABLE}. 
The latter is one of the lightest well known scalar isoscalar mesons.
The analysis of stability of the equations 
by means of  dimensional arguments
is also discussed in the section IV (Derrick's analysis).
In section V we present our final comments, also
 discussing some ways of improving and extending the model.

\section{ Scalar field coupling}

We do not describe the Skyrmion model at length since this is 
done extensively in the literature.
One natural way of imposing a coupling of a scalar field 
(such as the chiral partner of the pion) to the
skyrmion model is to consider the linear realization of the sigma model.
The degree of freedom of the sigma 
may be replaced by a variable (dynamical)
chiral radius, $\eta(r)$, which becomes the new
dynamical degree of freedom: 
$$ f_{\pi} \to  \eta = \sqrt{{\bf \pi}^2 +\sigma^2} = v + \epsilon,$$ 
where $v = f_{\pi}$ is the classical value of the scalar 
field and $\epsilon$ its fluctuation \cite{WEINMANE}.
However, the resulting coupled  model (roughly the linear sigma model) 
has no stable solitonic solution with non trivial topology
as discussed later in this paper.
Thus we  include, as it is  usually done in the standard 
Skyrmion model,  
 the fourth order stabilizing term for the pionic sector.
The modified Skyrmion Lagrangian density reads:
\be \label{2}
\displaystyle{ \bL = \frac{\eta(\br)^2}{4} Tr ( \partial_{\mu} U^{\dagger} 
\partial^{\mu} U ) + \frac{1}{32 e^2} Tr \left( \left[ U^{\dagger} 
\partial_{\mu} U , U^{\dagger} \partial_{\nu} U \right]^2 \right)  + 
\frac{1}{2} \mu^2 f_{\pi} \eta(\br) \left( Tr U + Tr U^{\dagger} -2 \right)
+ \bL_{\eta}  },
\ee
where the first term is the Lagrangian density of the 
 linear sigma model and the second one is introduced
in order to stabilize the soliton \cite{SKYRME}.
While the first term becomes scale invariant with the 
quadratic coupling,
the second one is scale invariant by itself and thus no such coupling 
has been considered in this work for this term 
\footnote{As a matter of 
fact, it is  possible to introduce a scale invariant 
coupling in such  term.}.
$\mu^2$ is the pion mass whose Lagrangian term breaks slightly the 
chiral symmetry and  scale invariance as well, at least in this form. 
$U$ is a chiral invariant matrix  parametrized in terms
of the profile function  $F(\br)$
within the Hedgehog ansatz (given by $\hat{\bpi} = \hat\br$):
$$U ( \br ) = \cos(F(\br)) + i { \btau \cdot \hat{\bpi}} \; \sin(F(\br)).$$
Finally, the scalar field self interacting Lagrangian is chosen 
such that one recovers a classical value in the vacuum
due to the (chiral) spontaneous symmetry breaking:
\be
\displaystyle{ \bL_{\eta} = \frac{1}{2} \partial_{\mu} \eta(\br) 
\partial^{\mu} \eta(\br) - \frac{1}{2} m^2_{\eta} \eta^2(\br) - 
\frac{1}{4}\lambda \eta^4(\br) }.
\ee
The scalar field mass  term breaks scale invariance but it is necessary
for the consistency of the model.
The bare scalar field mass parameter is  $m^2_{\eta}$, being related to the 
physical scalar 
field mass ($m_{\epsilon}^2$) by the following relation:
\be
\displaystyle{ m^2_{\eta} = \frac{1}{2} (\mu^2 - m^2_{\epsilon} ). }
\ee
The value of the condensate in the vacuum is found by the minimization of 
its potential.
As stated above
we consider two different possibilities for the physical particle associated
with the scalar field. 
The $\eta$ field is expected to correspond to the lightest 
scalar-isoscalar meson. 
It may be a very broad
scalar resonance in the s-wave of low energy $\pi-\pi$ scattering,
with a mass in the range $500-800 \;$ MeV that seems to have
been observed in other processes
\cite{MESTRADO,TORNQVIST,BEDIAGA}.
In this case it would be associated to the chiral partner of the pion  
eventually corresponding to a Higgs field for 
Hadronic interactions 
\footnote{ Its origin, on the other hand, 
lies on the chiral spontaneous symmetry 
breaking of QCD vacuum.}.
We consider two different values for the scalar-isoscalar 
particle mass in order to investigate its effects on 
nucleon properties.
The coupling constant $e$ may be related to the pion-pion scattering
coupling as done in \cite{TRUONG,DONOHOLS}, i.e.,
to a particular choice of the low energy constants (LECs) 
of some terms from
the Chiral Perturbation Theory expansion \cite{CBW02}.
In the works just quoted above the authors obtained: $e \simeq 4 -7$.
Large $N_c$ analysis with extrapolation to $N_c=3$ 
using sum rules of \cite{DOSCH} leads to the approximated values:
$e\simeq (7.6-12)/\sqrt{N_c}.$
These values are, in principle, independent of $f_{\pi}$.

The coupling to the quadratic Lagrangian term
of the Skyrmion model, as described above, is uniquely defined
by the linear implementation of chiral symmetry. 
It is scale invariant. 
It is consistent with the consideration that 
the chiral radius is directly and unambiguosly related to the quark-antiquark 
condensate of QCD and, besides, to the pion decay constant,
at least in the vacuum and  
at the tree level.  
For the coupling of the quartic stabilizing Skyrme
term to the scalar field we found no scale invariant form.
 
Along this work we basically assume 
the chiral radius to be a  dynamical variable, 
being equivalent to the sigma field. 
In this case one is led to the following 
replacement  for
the quadratic Skyrmion Lagrangian term:
\be \ba{ll} \label{subst}
\displaystyle{ U(\br) \to \alpha(\br) U(\br),\;\;\; \mbox{with}
\;\;\; \alpha (\br) = \frac{\eta(\br)}{f_{\pi}}. }
\ea
\ee 
However, this replacement in the fourth order Skyrme term
yields scale non-invariant Lagrangian terms and 
unstable solutions in the
physical region for the parameters of the model. 
The lack of stability was also 
checked by the Derrick's argument \cite{DERRICK}
discussed in section IV. 


Finally, one can notice 
that the fourth order Skyrme term may be derived
 from the coupling of the sigma model with the 
vector meson, $\vec\rho_{\mu}$, as a hidden or massive gauge boson 
\cite{TRUONG,REVIEW-VEC}.
The above prescription for the
function $U(\br)$ may yield a $\eta(\br)$-dependent $\rho$ mass and $e$ coupling
such that scale invariance is mantained.
This will be developed elsewhere.
Although this is a very interesting idea by itself, 
after the elimination of the 
$\rho$ degrees of freedom, the prescription seems to
lead to other derivative 
and complicated terms which also may break scale invariance. 
At last it is interesting to preserve scale invariance 
in spite of the massive terms.

\section{ Calculations}

In this section we investigate the solitonic solutions relative
to equations of the fields of the model and afterwards 
their consequences for static nucleon observables.
Adopting the usual hedgehog ansatz, 
$\hat{\bpi} = \hat{\br}$, the modified Lagrangian is written as:
\be
\displaystyle{ L = 4 \pi \int dr r^2 \left\{ -\frac{\eta^2}{2} F'^2 -
\frac{sin^2(F) }{e^2r^2} F'^2- \frac{sin^2(F)}{r^2} - \frac{sin^4(F)}
{2e^2r^4} + \mu^2 \eta f_{\pi} cos(F) + \bL_{\eta} \right\} , }
\ee
where $F = F(r) $,  $\eta = \eta(r)$ and $\; ' \;$ is the derivative
with relation to the radial coordinate $r$.

The Euler-Lagrange equations for this system with the 
change of variable $r \to r f_{\pi}$ are given by:
\be  \label{4} \ba{ll}
\displaystyle{ F''\left( r^2\eta^2 + \frac{ 2sin^2(F)}{e^2} \right) + 
2 r \eta^2 F' + \frac{sin(2F)}{e^2} F'^2 - \eta^2 sin(2F) -
\frac{ 4sin^3(F) cos(F)}{2e^2 r^2} \{+\} } \\
\displaystyle{ - \mu^2 \eta sin(F) r^2 f_{\pi} + 
2 r \eta \eta' F' = 0,  }
\ea
\ee
\be \label{4a}
\displaystyle{ \eta'' + \frac{2}{r} \eta' - \eta \left( F'^2 +
\frac{2 sin^2(F) }{r^2}  + m_{\eta}^2 \right) + 
\mu^2 cos(F)  f_{\pi}
- \lambda \eta^3 =0  .}
\ee
In the spatially homogeneous case of $\eta (r) = f_{\pi} $ (constant),
 the usual 
skyrmion equation and solutions are obtained
for the first equation.
That value for the sigma ``condensate'' (classical field) naturally
emerges as a boundary solution in the limit $r \to \infty$, 
i.e., in the vacuum, far from the topological soliton for 
the solutions of interest.
In the limit of 
 $m^2_{\eta} \to \infty$ the equations become
uncoupled, what is verified numerically.

For the scalar field equation we consider the following 
conditions: 
\be \ba{ll} \label{boundcondeta}
\displaystyle{ \eta(r\to \infty) = f_{\pi} , }\\
\displaystyle{ \eta'(r\to \infty) = \eta'(r = 0) = 0.}
\ea
\ee
In this limit, in the vacuum, a constraint between the 
scalar field parameters is obtained. 
By minimizing the (homogeneous) scalar field potential
 from equation (\ref{4a}) we obtain:
\be \label{lambda} \ba{ll}
\displaystyle{ \lambda = \frac{\mu^2 - m_{\eta}^2}{f_{\pi}^2}.}
\ea
\ee
The corresponding solutions for the spatially 
non homogeneous case are presented later.
Spontaneous chiral symmetry breaking occurs
as long as
\be \ba{ll} \label{condit}
\displaystyle{ \frac{\lambda}{m_{\eta}^2} < 0,}
\ea
\ee
which provides a condition for the existence of the 
``sigma condensate'' at the classical level.
Quantum fluctuations may considerably change this
 \cite{FLB,PRD2001}. 
They will be investigated elsewhere.

For the equation of the profile function $F(r)$ the usual boundary 
conditions still hold, i.e., winding number $n= 1$, for which one considers
$F(r=0) = \pi$ and $F(r\to \infty ) = 0$
\cite{ADKINS}. 
From expression (\ref{4a}) one sees that it is possible to define
an effective mass parameter for the field $\eta(r)$.
Namely:
\be \label{5a}
 m^2_{eff} = m^2_{\eta} + F'^2 + \frac{2sin^2(F)}{r^2}.
\ee
Far from the nucleon this effective mass is just equal to 
$m_{\eta}^2$, but
inside the skyrmion and close to it the contributions from the two 
other terms may be relevant.

\subsection{ Numerical results}

We are thus faced with two differential equations with 
two point-type boundary conditions. They are solved in
an iterative way by the ``shooting'' method using 
fourth order Runge Kutta with the boundary conditions shown above.

The numerical solutions for the skyrmion profile function
are presented in figure 1 and, in figure 2, the solutions for the 
condensate $\eta(r)$. 
Three different cases are shown
considering  $f_{\pi}=93 \;$ MeV.
The following  values for the other
input parameter were used for each  solution:
\be \ba{ll}
\displaystyle{ (1) \;\; m_{\epsilon} = 
980 \mbox{MeV} \;\;\; \mbox{and} \;\;\; e=4 ,}\\
\displaystyle{ (2) \;\; m_{\epsilon} = 
980 \mbox{MeV} \;\;\; \mbox{and} \;\;\;  e=5, }\\
\displaystyle{ (3) \;\; m_{\epsilon} = 
560 \mbox{MeV} \;\;\; \mbox{and} \;\;\; e=4.}
\ea
\ee
From figure 1 we notice that the skyrmion radius becomes a little smaller 
and that inside the topological soliton the scalar 
field tends to decrease, i.e., the nucleon ``digs a hole'' in 
the vacuum otherwise fulfilled with the condensate. 
For smaller (bigger) values of the scalar 
field mass the ``hole'' -- which is in fact akin to the ``bag''
of other models --
becomes  deeper and wider (shallower and narrower).
This is a mechanism for bag formation equal to those 
studied in \cite{GJJS86,WAWA,MEISS-SCHE}.
As a matter of fact, the Skyrmion model and the constituent quark 
model were shown to yield equivalent results in the large $N_c$ limit
\cite{RISKA}, thus the  bag structure should be expected in both models. 

In fact, we find that inside the topolotical soliton 
the effective mass given in expression (\ref{5a}) 
should be smaller in modulus, being even
 positive, thus violating  condition (\ref{condit})
for the occurrence of spontaneous symmetry breaking.
It can be seen as  the restoration of the
chiral symmetry inside the soliton (nucleon) for not high
values of $m_{\eta}^2$. 
Therefore there is a sort of a ``critical 
value'' for the scalar particle mass above which the 
chiral symmetry is restored inside the nucleon. 
This value depends on the other parameters of the model
and 
on the level of approximation for the quantum
fluctuations of the fields.

Although it is not clear whether it is physically 
meaningful we describe the following curious issue.
We have also investigated the possibility that at the origin, i.e.,
inside the nucleon, the classical scalar field is larger
than in the vacuum. 
For a very particular fine tuned value of $\eta(r \to 0)$ - which depends
on $e$, $m_{\eta}^2$ and $\eta´(r \to 0)$ - 
we found a solution which goes asymptoticaly to $f_{\pi}$, as 
required and expected. We have obtained this only for large mass values.
However this solution exhibits a sort of small amplitude oscillation,
which decreases progressively with r,
having large wavelength 
around the region of the topological soliton. 
It would correspond 
to a ``stronger chiral symmetry breaking'' inside the soliton.
We leave the investigation of such solution
for another work \cite{CONTINUE}.


Strong and electromagnetic properties usually obtained for the
skyrmion \cite{ADKINS}
 were calculated considering the above scalar field coupling.
The energy of the lightest baryons (nucleons) can be written as:
\be \label{7} \ba{ll}
\displaystyle{ M = 4\pi \int^{\infty}_0 d r r^2 
\left\{ \frac{\eta^2}{2} \left( F'^2 + \frac{2 sin^2(F)}{r^2} \right)
+ (1-cos(F)) \mu^2 \eta f_{\pi} + \frac{\eta'^2}{2} + \frac{m^2_{\eta}}{2} \eta^2
+ \frac{\lambda}{4} \eta^4 + \right. }\\
\displaystyle{ \left. + \frac{1}{e^2} \left( 
\frac{sin^2(F) F'^2}{r^2} + \frac{sin^4(F)}{2r^4} \right) 
+ \frac{1}{ {\cal I}}
- E_{vac}  \right\}  ,} 
\ea
\ee
where ${\cal I}$ is the 
moment of inertia \cite{ADKINS} which now depends on $\eta(r)$
and $E_{vac}$ is the vacuum energy
due to the dynamical chiral symmetry breaking
to be subtracted:
\be
 E_{vac} = \left( \frac{m_{\eta}^2}{4} - \frac{3 \mu^2}{4} \right).
\ee

In table 1 we compare the values of observables obtained for the 
usual Skyrmion model, our model (for the three cases shown in 
figures 1 and 2) and experimental data. 
A higher value for $e$ ($e=5$) yield better
 values for the mass spectrum as it is well known. 
However almost all the other
observables (mainly the electromagnetic properties) 
still present values in disagreement with the experimental ones.
On the other hand, considering smaller values of the scalar field mass,
$m_{\epsilon} = 560 \; MeV$, 
a better agreement with the experimental values for most  observables
is obtained. 
The most remarkable
exception is the behaviour of the axial coupling 
constant $g_A$. 
The $g_A$ coupling is calculated in two ways:
1) in the limit of zero momentum of the nucleon axial form factor
and 2) by means of the Goldberger-Treiman relation  \cite{ADKINS}
in which case all the parameters can be calculated from 
the resulting Skyrmion solutions. 
For $g_A$ and $g_{\pi NN}$ a smaller 
value of $e$ would be desirable, what would, in principle, make the values 
of the other observables to be in 
disagreement with  experimental ones.

\section{ Derrick's analysis of stability }

We have also performed the Derrick's analysis in order to verify
the stability of the ``finite sized" solutions of the 
differential equations \cite{DERRICK}. 
It consists in checking, by dimensional arguments (scale transformations),
 whether there is a scale in which the total energy 
of the system is finite yielding stable finite size
solutions for the corresponding equations. 
One has to consider scale transformations for the
spatial coordinate variable $$r \to r' = \frac{r}{\alpha}$$ and for the 
dimensionful field 
$$\eta (r) \to \beta \eta(r').$$ 
In fact it is more suitable to do it for the shifted equivalent field: 
$\theta(r) = f_{\pi}- \eta(r)$.
Under  the above transformations we verify whether the energy density
has a stable minimum with relation to the scale transformation 
parameters, i.e.:
\be \label{10a} \ba{ll}
\displaystyle{ \frac{\delta M}{\delta \alpha} =0 ,\;\;\;\;\;\;\;\;\;\;
 \frac{\delta M}{\delta \beta} =0  , }
\ea
\ee
\be \label{10b} \ba{ll}
\displaystyle{ \frac{\delta^2 M}{\delta \alpha^2} > 0, \;\;\;\;\;\;\;\;\;\;
 \frac{\delta^2 M}{\delta \beta^2} > 0 .}
\ea
\ee

This analysis was performed for the following cases:

\noindent (i) Only the sigma model with the scalar field which has
self interacting potential, i.e., no fourth order term for the pion
field in the Lagrangian density (\ref{2});

\noindent 
(ii) The inclusion of the usual quartic Skyrme term exactly as it stands in 
expression (\ref{2});

\noindent
(iii) The coupling of the scalar field in the fourth order Skyrme
term, breaking the scale invariance, with the ansatz 
of expression (\ref{subst}).
This modification yields, respectively, 
the following modified and additional Lagrangian
density terms:
\be
\displaystyle{ {\cal L}_4´ =  \frac{1}{f_{\pi}^4 e^2} \left\{
\eta^4(r) \left( \frac{s^2 F'^2}{r^2} + \frac{s^4}{2r^4} \right)
+ \frac{\eta^2 \eta'^2 s^2}{4 r^2} \right\}  .}
\ee

\vs

In the first case (i)
there are no stable solutions for the skyrmion
profile function ($\delta^2 M/\delta \alpha^2 < 0$). 
It becomes clear
once again the importance of the higher order derivative terms 
in order to keep the soliton against
collapse,
 in spite of the problems with the convergence of the
 derivative series \cite{AITCHFRA}.

\vs

Case (ii) is stable provided that the chiral symmetry breaking term
is small (as always assumed) 
compared to the self interacting scalar field term as it can 
be seen from the second equation of (\ref{IIalpha}) below.
The conditions for stability are given by:
\be \ba{ll}  \label{IIalpha}
\displaystyle{
\frac{\delta^2 M}{ \delta \alpha^2} > 0  \to \int^{\infty}_0 r^2 dr
\frac{1}{e f_{\pi}} 
\left[ 2  \left(2 sin^2(F) \frac{(F')^2}{r^2} +
\frac{sin^4(F)}{r^4} \right) - \frac{\eta^2}{4} 
\left( (F')^2 + \frac{2 sin^2(F)}{r^2} \right) - 
\frac{1}{4} (\eta')^2 \right] >0,  
}\\
\displaystyle{
\frac{\delta^2 M}{ \delta \beta^2} > 0 \to  \int^{\infty}_0 r^2 dr 
\left( \lambda \eta^4 + \mu^2\eta f_{\pi}(cos(F)-1) \right)
> 0 
.} 
\ea
\ee
The first condition does not depend on $e$ neither on $f_{\pi}$.
However the second one depends on the value of $\lambda$.  
This is found in numerical calculations where for very small values of 
$\lambda$ (and therefore of $m_{\eta}$)
the solution for $\eta(r)$ becomes so stiffer that it makes 
the numerical calculations more unreliable and the solutions eventually
inexistent for the boundary conditions we consider.

\vs

For  case (iii) we have the following stability conditions:
\be \ba{ll}  \label{IIIalpha}
\displaystyle{ 
\frac{\partial^2 M }{\partial \alpha^2} > 0 \to
 \int^{\infty}_0 r^2 dr \left\{ 
\frac{4}{f_{\pi}^2} \left( 2 sin^2(F) \frac{(F')^2}{r^2} + 
\frac{sin^4(F)}{r^4} \right) \eta^4 - 
\frac{\eta^2}{4} \left( (F')^2 + \frac{2 sin^2(F)}{r^2} \right)
- \frac{(\eta')^2}{4} \right\} >0,
}\\
\displaystyle{ 
\frac{\delta^2 M}{ \delta \beta^2} > 0 \to
 \int^{\infty}_0 r^2 dr 
\left(2 \lambda \eta^4 + 16 e^2 \eta^4 \left( \frac{sin^2(F) (F')^2}{r^2}
+ \frac{sin^4(F)}{2 r^2} \right) - \mu^2 f_{\pi} \eta (1 - cos(F))
\right) > 0 .
} 
\ea
\ee
The first of these equations depends on a scale ($f_{\pi}$ or $e$) for 
the stability  and the second one depends
on the relation between two parameters, $\lambda$ and $e$. 
This can be seen by means of an
usual change of variable such as $r \to e f_{\pi} r$.

With this analysis we see that the coupling of the scalar field to the
fourth order Skyrme term as in prescription (\ref{subst}) is
not suitable for the stability of the Skyrmion solution as 
it was argued before.

\section{ Summary and Conclusions}

We have investigated the effect of coupling a classical
scalar isoscalar 
field to the usual Skyrmion model as a degree of freedom 
whose corresponding quantum would be the chiral partner of the pion.
The dependence of the observables on the values of the 
scalar field mass and on the 
fourth order coupling term $e$ were also considered.
Smaller values of the $\sigma$ mass favor better agreement
for most of the nucleon static properties, except $g_A$.
Besides the expected solution of small values of the scalar condensate
inside the nucleon, 
we found the possibility of a higher value
for the classical field (and also for its mass) 
inside the nucleon, but for restricted values
of the free parameters. 
This would be equivalent to a 
further,  stronger, chiral symmetry breaking
\cite{PRD2001,QMC-Panda}.
This feature is not yet well understood and is being 
investigated \cite{CONTINUE}.

This non trivial solution for the chiral radius (equivalently,
for the sigma classical field) may suggest 
that the chiral symmetry is realized in the Wigner-Weyl
mode inside the soliton instead of the Nambu-Goldstone mode, which 
is expected to hold in the vacuum. 
The scalar field mass is another free parameter that determines
the form of the potential
which presents (or not) the spontaneous symmetry breaking.
The coupling to the nucleon allows the definition of
an  effective mass parameter, given  by expression (\ref{5a}),
which may be even 
positive inside the nucleon for the solutions shown
in the figures while outside (in the vacuum)
it becomes negative yielding the ``mexican hat''-type potential
for $\eta$.

In this work the pion mass term was considered 
to break explicitely
chiral symmetry and to have the more standard form. 
However  this form is not fully
determined unless for the requirement of yielding the leading terms 
of the axial current divergence - PCAC.
Furthermore its coupling to the scalar field is not at all
uniquely determined and it may be expected to be 
related to the trace anomaly of QCD \cite{BEANE-BIRA}
and  certainly to the chiral SSB.
Alternative schemes for 
that will be shown elsewhere.

As discussed above there is a procedure for obtaining the fourth order 
stabilizing term from a coupling to the $\rho$ meson field (and also 
to its axial chiral partner $A_1$).
It is based on the elimination of the vector meson degrees of freedom
by performing the infinite mass limit and keeping the
KSRF relation \cite{REVIEW-VEC}: 
$m^2_{\rho} = 2 g_{\rho \pi \pi}^2 f_{\pi}^2$.
However, we have consistently considered that inside the nucleon
 $f_{\pi}$ becomes a dynamical degree of freedom whose value
may be seemingly reduced. 
Therefore the limit
of large rho mass is not necessarily reliable anymore, thus imposing
difficulties in order to relate one picture (with the $\rho$)
to another (without it). 
This will be shown in detail elsewhere.

\vs

{\bf \Large \bf Acknowledgements}

This work was supported by FAPESP, Brazil. 
The authors thank Dr. M. R. Robilotta for several discussions including the 
idea of prescription (\ref{subst}). I. P. C. also thanks Dr. F. F. da
Souza Cruz Filho for several fruitful discussions, which have taken
place in the Nuclear Theory Department of the University of
Washington, Seattle, U.S.A. 
This author is thankful to Dr. Larry
Wilets for their hospitality.  
F.L.B. thanks Dr. 
F.S. Navarra for motivating support and interest in the subject and 
interesting discussion.

\vs

{\bf Figure captions }

\vs

{\bf Figure 1.}

{\it Profile function F(r) for uncoupled skyrmion 
(solid line) and taking into account the coupling
to the condensate with input parameters: e=4 and 
$m_{\epsilon}=980 \;$ MeV (dashed line), e=5 and $m_{\epsilon}= 980 \;$MeV 
(dotted-dashed line) and e=4 with $m_{\epsilon}=560 \;$MeV (dotted line).
 }

\vs

{\bf Figure 2.}

{\it Condensate  $\eta(r)$ in the presence of 
the soliton for the three cases with coupling of figure 1. }

\vspace{1cm}

\begin{figure}[htb]
\epsfig{figure=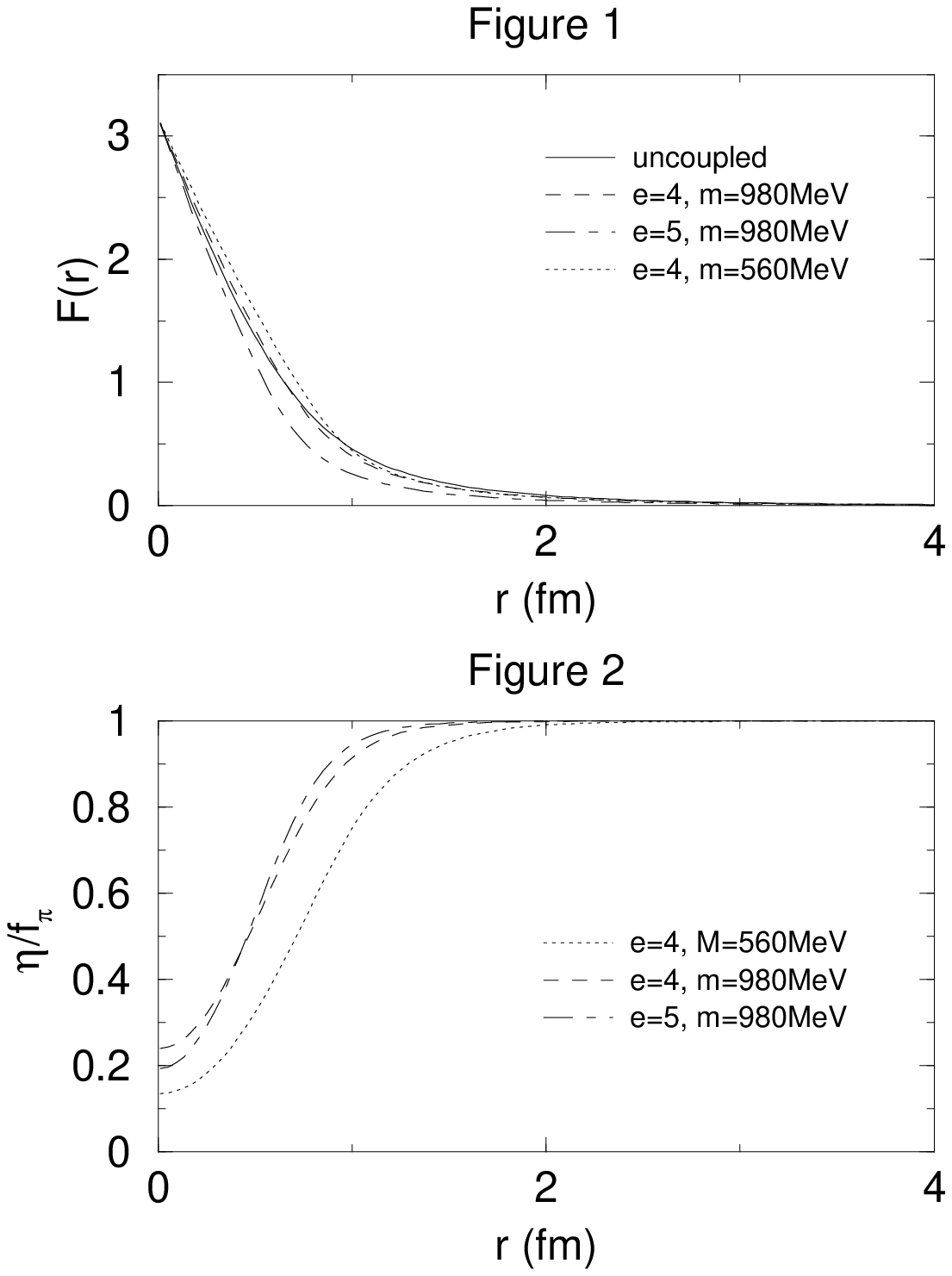,width=16cm} 
\end{figure}

\begin{table}[t]
{\footnotesize \caption{ Values of the observables calculated for the
free skyrmion model and the other three cases discussed in the text 
(with $f_{\pi} = 93 \; $ MeV). 
$g_A$ is calculated in two ways, from 
the zero momentum nucleon axial form factor as well as from 
the Goldberger--Treiman relation.
} }
\centerline{\begin{tabular}{c c c c c c c c}  \\ \hline \hline
 observable  & free  & e=4, & e=5, & e=4,  & experiments \\
 & skyrmion & $m_{\epsilon}=980$MeV & $m_{\epsilon}=980$MeV &
  $m_{\epsilon}=560$ MeV & \\
 \hline
$ M_N      $  (MeV) & 1819 &1578 &1282 & 1436  & 939     \\
$M_{\Delta} $(MeV)  & 2069 &1876 &1830 & 1722 & 1210     \\
$\lambda$ (MeV$^{-1}$)& 166.7 & 198 &366 & 191 & ---    \\
$\sigma$-term     &49.6 &62 & 51& 54 & 50 $\pm$ 20      \\
 $<r_E^2>^{\frac{1}{2}}_B $(fm) &.51 & .51 & .43 &.56   &.72     \\
$<r_E^2>^{\frac{1}{2}}_V $(fm)  &.87 & .80 &.71 &.82   &.88     \\
$<r_M^2>^{\frac{1}{2}}_B $ (fm) &.82 & .53 & .44 &.56   &.56     \\
$<r_M^2>^{\frac{1}{2}}_V $ (fm) &.80 & .62 & .55&.67   &.64     \\
$<r_E^2>^{\frac{1}{2}}_p $ (fm) &.86 & .67 &.59 &.71   &.70     \\
$<r_E^2>^{\frac{1}{2}}_n $ (fm) &-.24 &-.19 &-.16 & -.18  &-.12     \\
$<r_M^2>^{\frac{1}{2}}_p $ (fm) &.66  &.61 & .53 &.63   &.86$\pm$.06  \\
$<r_M^2>^{\frac{1}{2}}_n $ (fm) &.68  & .64 & .59& .65   &.88$\pm$.07  \\
$\mu^B (magn.) $  &   .34    & .35 & .37  & .36 & .44        \\
$\mu^V (magn.) $  &  3.64    & 2.66 & 1.17 & 2.5 & 2.35        \\
$\mu_p (magn.) $  & 3.97    & 3.01 &1.54  & 2.9 & 2.79     \\
$\mu_n (magn.) $  &  -3.30  & -2.31 & -.80   &-2.1 & -1.91     \\
$\mu_{\Delta++} (magn.) $ &7.55 & 5.84 &3.21 & 5.6   &4.7- 6.7      \\
$\mu_{N\Delta} (magn.)$  & 2.66 & 2.24 & 1.21 & 2.32  & 3.3     \\
$g_A $ ($g_A$ GT rel.) & 1.01(1.01)& .76(0.84) &.49(.54)&.78(.83)& 1.23 \\
$g_{\pi \pi N}$  & 20.3 &14.5 &7.5 & 12.7 & 13.5      \\
$<r_A^2>^{\frac{1}{2}} $ fm &.49 & .86 & .85 & .48  & .68    \\
$<r_{\pi}^2>^{\frac{1}{2}} $fm  & .6 &.50 & .37 & .28  & .54    \\
\hline \hline    
\end{tabular} }
\end{table}

\end{document}